\documentclass[a4paper]{article}
\usepackage[ansinew]{inputenc}
\usepackage{times}
\usepackage[T1]{fontenc}
\usepackage{graphicx}
\usepackage{geometry}
\usepackage{amssymb}
\usepackage{amsmath}

\usepackage{rotating}

\usepackage{xcolor}
\colorlet{shadecolor}{gray!25}
\usepackage{subfig}
\usepackage{float}

\author{\normalsize Ludwig A. Hothorn,\\ 
\footnotesize Im Grund 12, D-31867 Lauenau, Germany (e-mail:ludwig@hothorn.de)\\ \scriptsize(retired from Leibniz University Hannover)}

\title{Simultaneous confidence intervals for the interpretation of primary and secondary effects in factorial designs\\ without a pre-test on interaction}
\begin{document}

\maketitle
\begin{abstract}
The analysis of low dimensional factorial designs with possible interactions is a relevant issue. Instead of the common pre-tests for interaction, a simultaneous inference procedure of the primary factor at the respective level of the secondary factor as well as pooled over all its levels is proposed.
\end{abstract}

\section{Introduction}\label{sec1}
In biomedical trials, the methodologically so predominantly focused one-way layouts already exist, but factorial designs with  commonly a few factors are the standard in the recent literature, particularly in biological experiments \cite{zhu2019}. From this it follows how to analyze the main effects considering the interaction effects appropriately?

\section{Model}
Here, we consider a generally unbalanced (i.e. replicated) factorial design with a few factors, e.g., 2 or 3 factors, where at least one factor has more than just 2 factor levels (to exceed the trivial 2x2 design). Notice, we consider qualitative factors only; the relationship between a quantitative covariate and a qualitative factor is a different issues, see e.g. \cite{Schaarschmidt2017}.  Without limiting generalizability, a scenario with only two factors is simplified below. From the point of view of interpretation, one factor $A_i$ is usually of primary interest, usually the one with $k>2$ levels, while the secondary factor $B_j$ should describe different conditions for the primary factor, such as males and females in randomized clinical trials \cite{Biesheuvel2002}.\\

A central inference issue hereby is the possible $A\text{x}B_{(ij)}$ interaction, but not in the sense of an existing or negligible interaction. In fact, it is usual to perform the inference between the levels $A_i$ either separately for each level $B_j$ (exactly if the global $F_{A\text{x}B}$-test is significant) or pooled over the levels $B_j$ (exactly if the global $F_{A\text{x}B}$-test is non-significant) by means of an ANOVA $F_{A\text{x}B}$-pretest (at whatever level) \cite{Cheung1996}. This approach contains several pitfalls, at least: i) the combination of pre- and main test may contain inconsistencies of the decision when controlling of the $\alpha$ level \cite{Kluxen2020}, ii) the different power dependencies of pre- and main test on the same $n_i$ (where commonly lower power for the interaction pre-test), iii) although already the consequence of a significant level $\alpha$-test on the separate analysis is not too clear, the test for negligible interaction should be performed as an equivalence test, which however itself depends on the tolerable threshold, the choice of which is unclear with respect to separate/pooled analysis \cite{Chen2007, Cribbie2016}, and iv)  the $A\text{x}B_{(ij)}$ interaction is a global test, which is sensitive to at least one, any interaction $A\text{x}B_{(ij)}$, but does not allow to say which one it is. In a not too rare data situation, however, this relevant interaction may occur only in one  (or a few) constellation(s), which is/are unimportant from the point of view of the primary factor. Then its power would be significantly reduced by a separate analysis on the $B_j$ levels, which is actually unnecessary for the selected $A_i$ levels of interest. I.e., a global interaction F-test is not the method of choice but interaction contrasts are more appropriate \cite{Kitsche2015}, which allow both global and individual statements jointly.\\
Mindful of these drawbacks, we propose an approach without any interaction pre-test: the simultaneous analysis between the levels of the primary factor for both per $B_j$ levels and their pooled levels. I.e., simultaneous in the sense of FWER control by means of multiple contrast tests \cite{Hothorn2006}. This allows the direct interpretation of the main effects of the primary factor, given existing or not existing interactions  as well of those of the secondary factor within a single-step approach. Obviously, there are two drawbacks to this approach: i) it is conservative compared to unadjusted ANOVA tests (but see the multiplicity-adjusted versions \cite{LH2022t}), ii) it defers the pre-main test problem to the interpretation of a large number of individual comparisons. It is thus limited to small dimensions of I, especially J (which is also recommended from a power point of view).\\
A similar approach is slicing into individual contrasts \cite{wludyka2015}, which however is oriented towards the analysis of a resulting cell means model using analysis of means (ANOM) \cite{pallmann2016}.

\section{Simultaneous confidence intervals  between the levels of the primary factor for both separate and joint secondary factor levels}
Instead of a pre-test approach, a simultaneous single-step procedure, considering of $\vartheta$ multiple contrast tests, e.g. Dunnett  tests \cite{DUNNETT1955} jointly, each of the $\vartheta$ levels of the secondary factor \cite{Hothorn2020t}. Extending this approach, the global Dunnett test and the $\vartheta$ secondary level tests are recommended simultaneously, since a problem-adequate interpretation is possible and the inherent loss of power is small due to its high correlation \cite{Hothorn2020h, LH2022t}.\\
First, a multi-way ANOVA model without interactions is formulated in the appropriate linear , generalized linear or mixed effect model according to the specific type of endpoint and the chosen design. Second, for the primary factor, the multiple comparison procedure appropriate to the design and experimental question is chosen as the multiple contrast test, e.g., comparisons to a control \cite{DUNNETT1955}, comparisons to the overall mean \cite{pallmann2016}, or all-pair comparisons \cite{Tukey1953}. These comparisons are formulated pooled across the levels of the secondary factor as well as separately for each level of the secondary factor.  Assuming for example, $A_i \;\text{where} \; i=0,1,...,k$ and $B_j \; \text{with}\; j=1,2$ Dunnett-type comparisons result in $ (2+1)k$ individual comparisons. Third, the simultaneous confidence intervals (or adjusted p-values) are estimated for this purpose by formulating their joint contrast matrix for this complex, user-specific contrast test \cite{Hothorn2020t}.  Technically, this can be realized relatively easily with the CRAN package multcomp \cite{Hothorn2008}.

\subsection{A parametric approach}
\begin{figure}[H]
	 \centering
		 \includegraphics[width=0.4\textwidth]{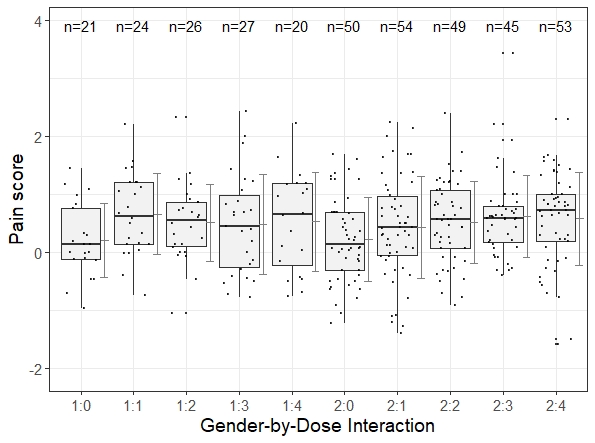}
	 \caption{RCT with Dose (0,1,2,3,4) in males (1) and females(2)}
	 \label{fig:EgbertBoxplot}
 \end{figure}

A first example considers the baseline adjusted abdominal pain score as primary endpoint in a unbalanced randomized dose finding trial with 4 doses and zero-dose placebo (primary factor) to treat the irritable bowel syndrome in males and females, as the secondary factor \cite{Biesheuvel2002} (see the box plots in Figure \ref{fig:EgbertBoxplot}.
 
This example assuming normal distributed errors was selected because the global interaction ANOVA F-test is not significant ($p=0.72$), and therefore, for the most part, only a Dunnett test would be performed across males and females. Table \ref{tab:Egbert} shows the one-sided adjusted p-values for the 12 individual comparisons for the joint test considering males, females and pooled data assuming homogeneous variances.  Significant increases of the endpoint with higher doses (2,3,4) vs. placebo in the pooled analysis are remarkable. On the other hand, a higher sensitivity in females compared to males of this dosing effect can be seen in their simultaneous analysis. The price of conservatism can be simply seen in this particular example: the analysis of the pooled data alone yields, for example, a p-value of 0.01 for the $[4-0]$ comparison compared with $0.026$ in the joint test here.

\begin{table}[ht]
\centering\small
\begin{tabular}{l|l|ll}
  \hline
 No& Type & Comparison & adjusted p-value \\ 
  \hline
1 & Males  &1:1 - 0 & 0.082 \\ 
  2 & &1:2 - 0 & 0.302 \\ 
  3 & &1:3 - 0 & 0.476 \\ 
  4 & &1:4 - 0 & 0.419 \\ \hline
  5 & Females&2:1 - 0 & 0.451 \\  
  6 & &2:2 - 0 & 0.146 \\ 
  7 & &2:3 - 0 & 0.030 \\ 
  8 & &2:4 - 0 & 0.070 \\ \hline
  9 & Pooled &p: 1 - 0 & 0.085 \\ 
  10 & & p: 2 - 0 & 0.041 \\ 
  11 & & p: 3 - 0 & 0.019 \\ 
  12 & & p: 4 - 0 & 0.026 \\ 
   \hline
\end{tabular}
\caption{Joint Dunnett tests for males, females and pooled data in the IBS trial}
\label{tab:Egbert} 
\end{table}
 
Of course, the interpretability of the joint test depends on only a few levels of the primary factor, and especially on only a few levels of the secondary factor, with a total of about two secondary factors being the highest. This is reasonably the case in many practical designs in bio-medicine. \\

A second unbalanced small $n_i$ example  in molecular plant science with a significant ANOVA concentration-by-type interaction F-test (p=0.02) was selected, with a primary factor dose on three ordered concentrations (0,1,10) and a secondary factor type on 2 levels  (wild type Col and mutant) \cite{zhu2019}. 

\begin{figure}
	\centering
	\includegraphics[width=0.3\textwidth]{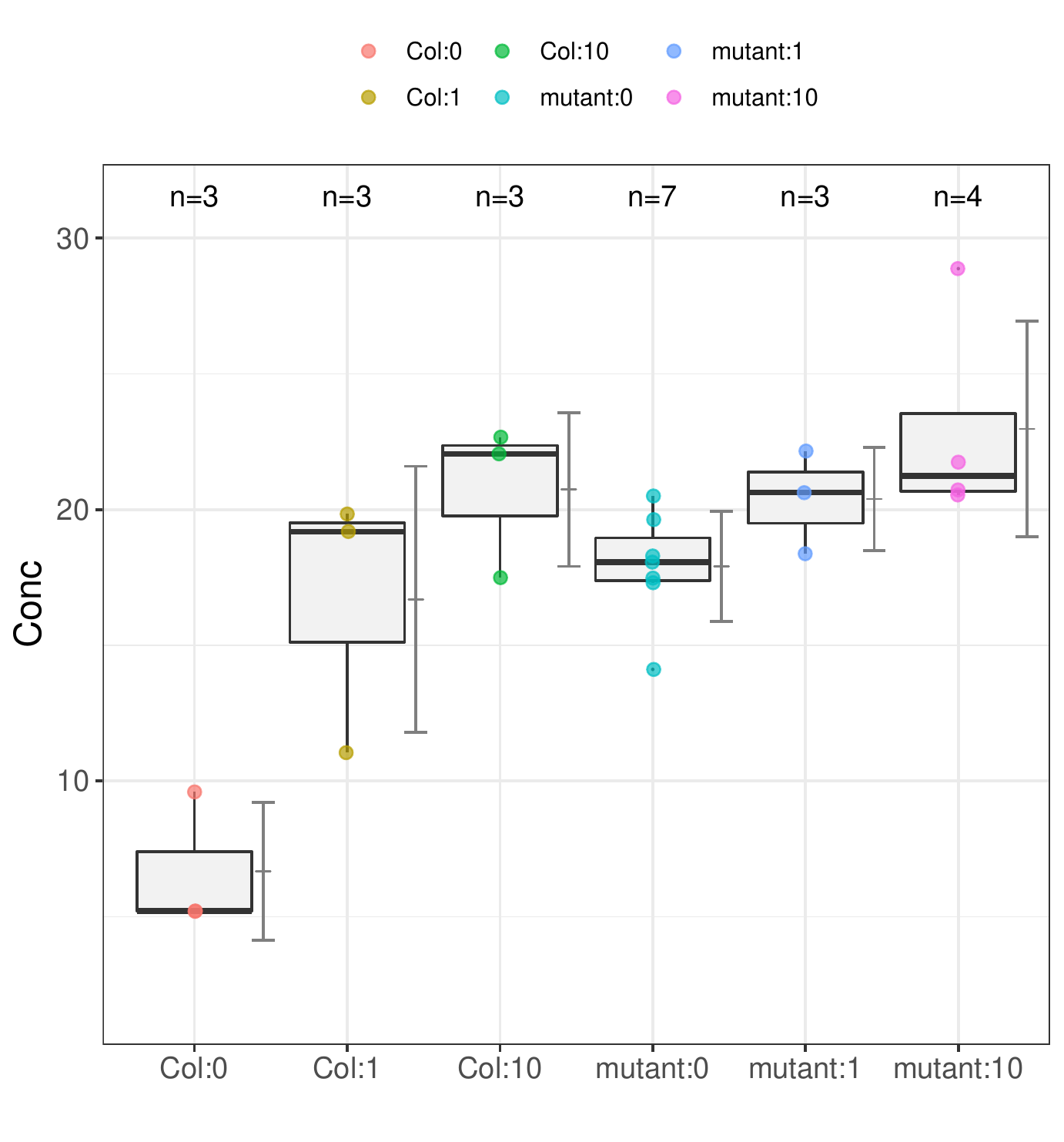}
		\includegraphics[width=0.3\textwidth]{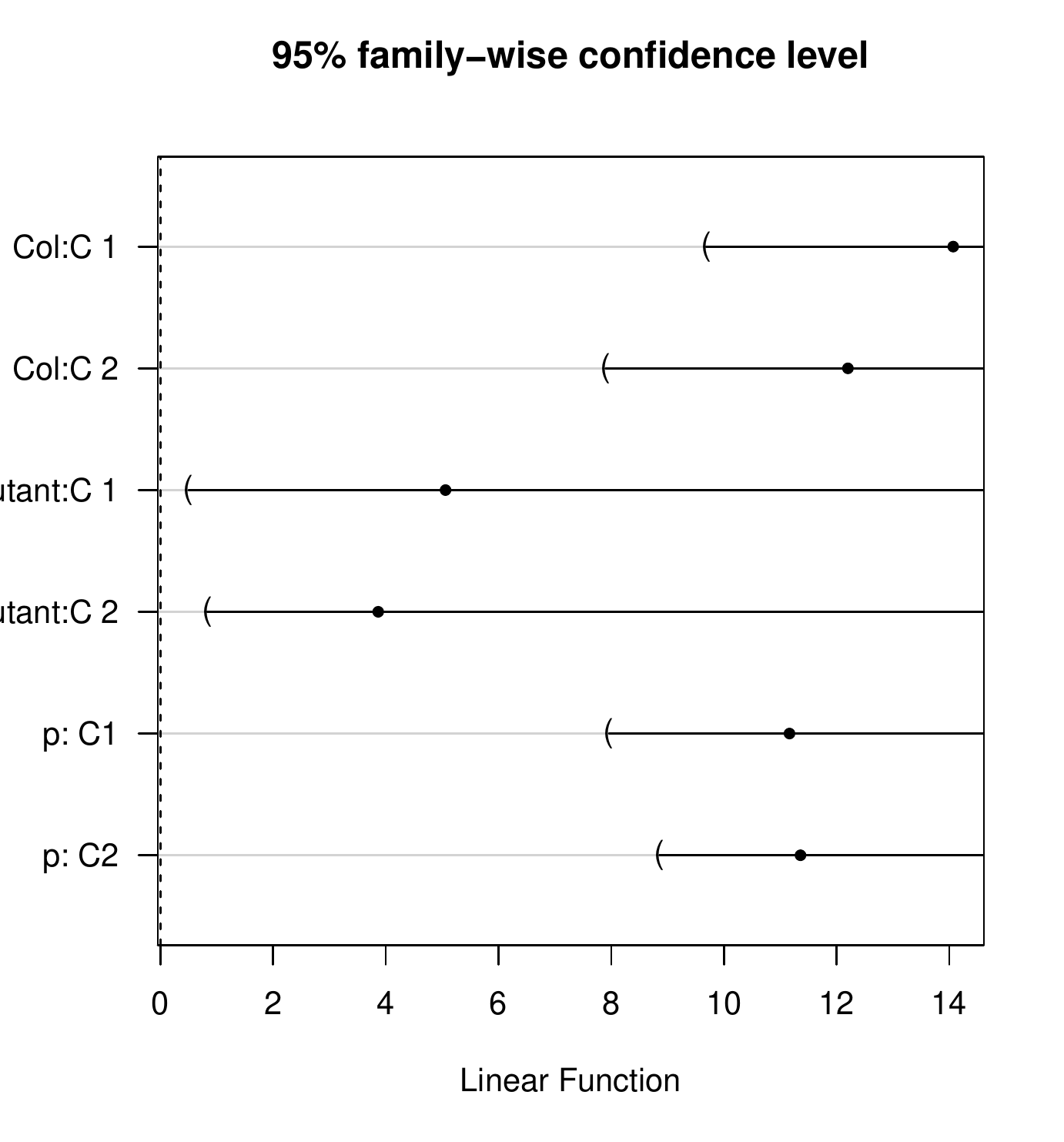}
		\includegraphics[width=0.3\textwidth]{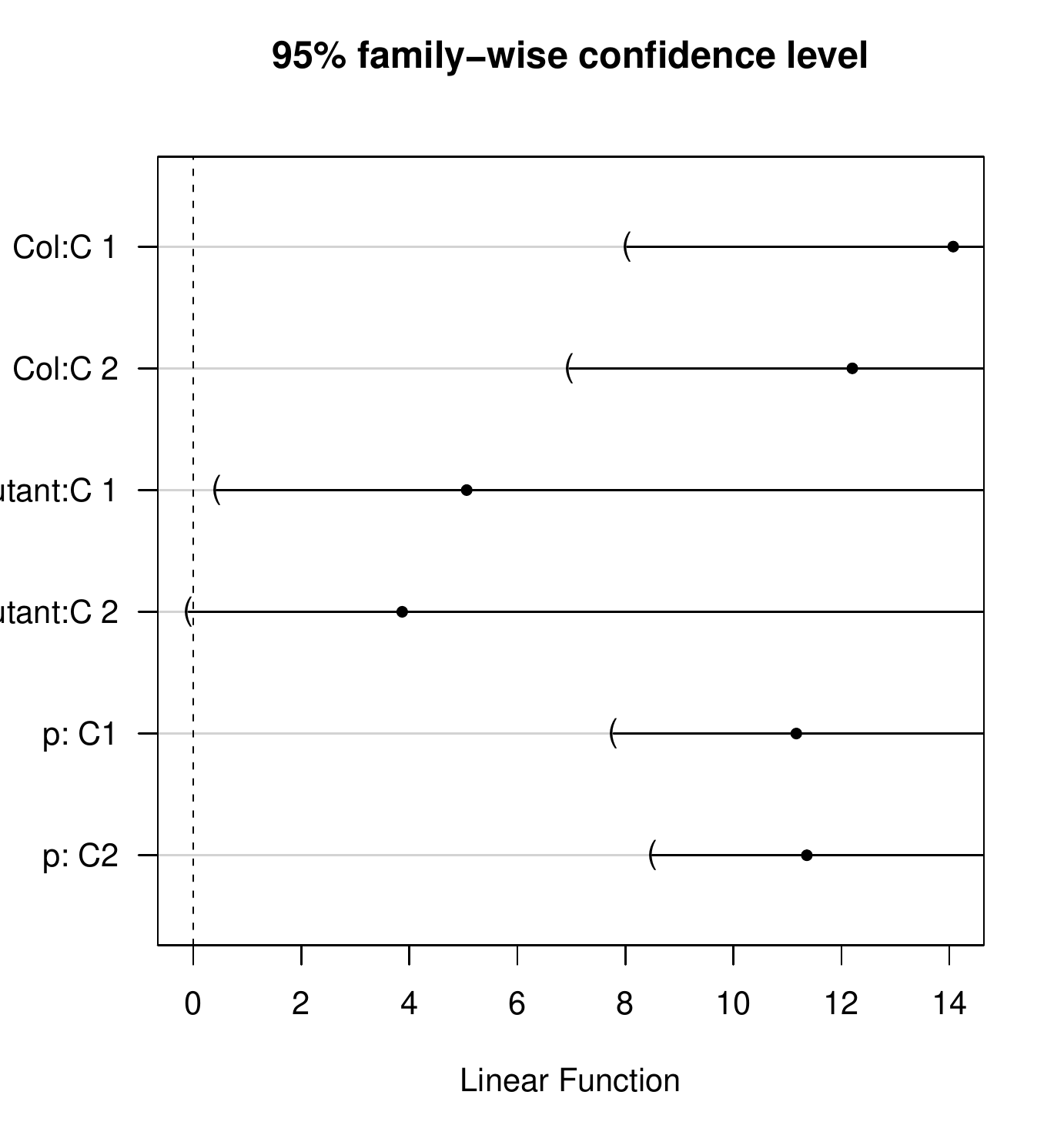}
	\caption{Boxplots and simultaneous Williams-type confidence intervals (middle panel: sandwich estimator)}
	\label{fig:X16}
\end{figure}

The box plots in Figure \ref{fig:X16} reveals a stronger dose-response relationship in the wild type compared to the mutant. 
Figure \ref{fig:X16} shows the simultaneous confidence intervals for Williams-type multiple contrast test \cite{WILLIAMS1971, Bretz2006} assuming a dose-response trend assuming either variance homogeneity or allowing heteroscedasticity by using a sandwich estimator \cite{Herberich2010}).

\begin{table}[ht]
\centering
\begin{tabular}{lllrrrr}
  \hline
Modeling & Contrast & Style & estimate & se & t & p \\ 
  \hline
heteroscedastic & 1-0  & per col & 14.0694 & 1.7904 & 7.8581 & 0.0000 \\ 
   & (10+1)/2-0 &  & 12.2018 & 1.7554 & 6.9509 & 0.0000 \\ 
   & 1-0 & per mutant & 5.0612 & 1.8620 & 2.7181 & 0.0290 \\ 
   & (10+1)/2-0 &  & 3.8667 & 1.2396 & 3.1194 & 0.0132 \\ 
   & 1-0 & pooled & 11.1644 & 1.3077 & 8.5377 & 0.0000 \\ 
   & (10+1)/2-0 &  & 11.3594 & 1.0219 & 11.1162 & 0.0000 \\ \hline
  homoscedastic & 1-0 & per col & 14.0694 & 2.4760 & 5.6823 & 0.0000 \\  
   & (10+1)/2-0 &  & 12.2018 & 2.1464 & 5.6847 & 0.0001 \\ 
   & 1-0 & per mutant & 5.0612 & 1.9007 & 2.6628 & 0.0329 \\ 
   & (10+1)/2-0 &  & 3.8667 & 1.6227 & 2.3828 & 0.0559 \\ 
   & 1-0 & pooled & 11.1644 & 1.3881 & 8.0430 & 0.0000 \\ 
   & (10+1)/2-0 &  & 11.3594 & 1.1745 & 9.6718 & 0.0000 \\  \hline \hline
	
	separate & 1-0 & col alone & 14.0694 & 2.9229 & 4.8134 & 0.0022 \\ 
   & (10+1)/2-0 &  & 12.0461 & 2.5314 & 4.7587 & 0.0023 \\ 
	 \hline
\end{tabular}
\caption{Williams-type adjusted estimates} 
\label{tab:X16}
\end{table}
Table \ref{tab:X16} reveals the largest value of the test statistics (and hence smallest p-value) for the pooled approach allowing heteroscedasticity for the Williams-type contrast $(10+1)/2-0$, i.e. a plateau shaped dose-response relationship is likely. This is because the smallest variance estimator in this particular case. The commonly-used separate analysis (after significant interaction) for the Col-stratum alone reveals a much larger p-value because of the much higher variance estimator, simply because of the reduced sample sizes, particularly relevant in such a small $n_i$ study.\\

A third example was selected with even an  qualitative interaction, i.e. the primary effect is not uni-directional in all levels of the secondary factor. 
\begin{figure}[H]
	\centering
		\includegraphics[width=0.50\textwidth]{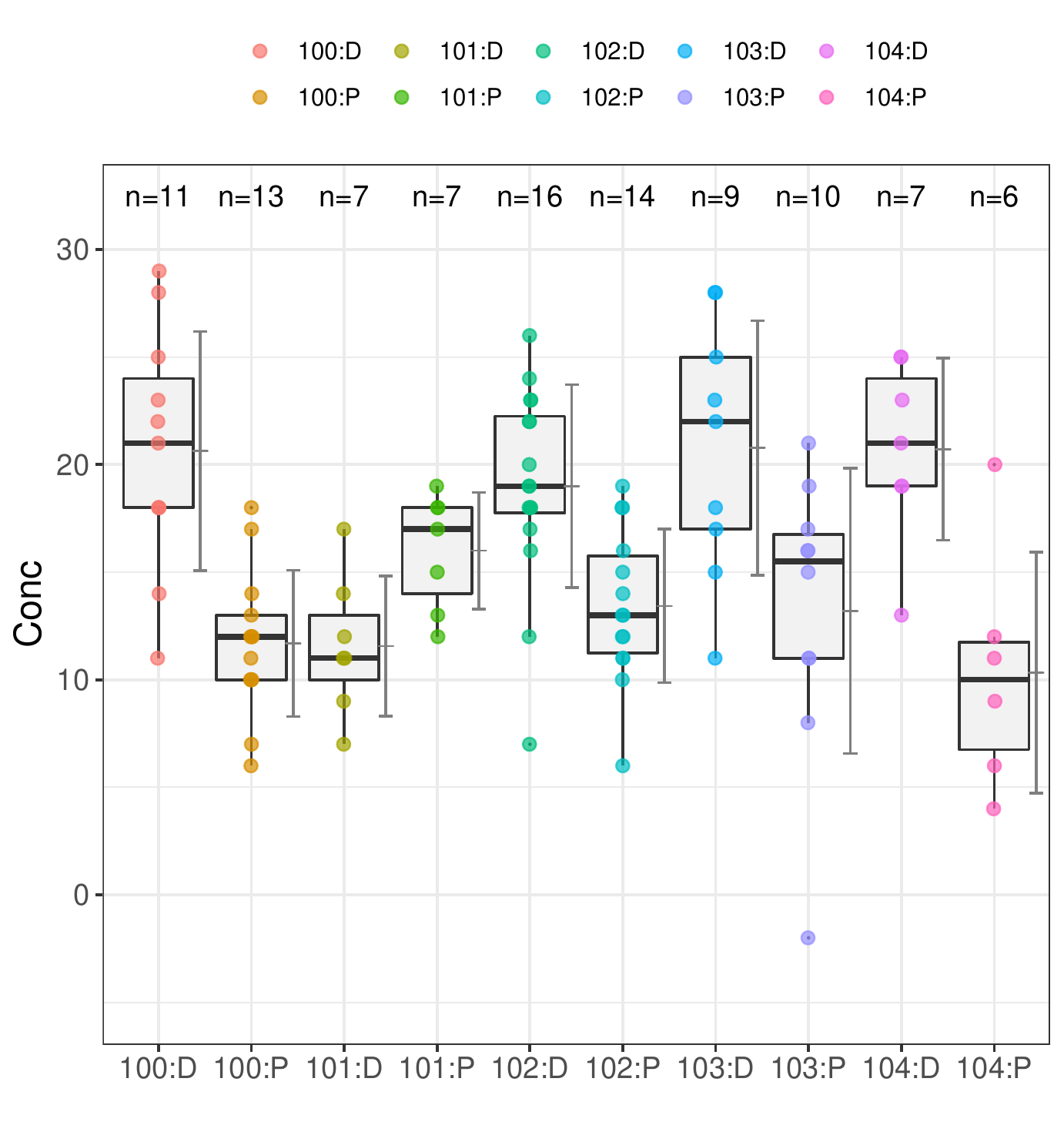}
	\caption{Boxplot multicentre depression trial}
	\label{fig:Dep}
\end{figure}

In this unbalanced small $n_i$, placebo-controlled, two-arm, randomized multi-centre depression trial the primary endpoint (change from the
baseline to the end of the 9-week acute treatment phase in the 17-item Hamilton depression rating scale HAMD17 score) \cite{Dmitrienko2005}.) reveals in centre 101 curious lower scores in the placebo group, see Figure \ref{fig:Dep}. The separate, un-adjusted per-centre p-values for an decrease are $(0.000045, 0.994, 0.00036,0.0064,0.00095)$

\subsection{A nonparametric version}
Multiple contrasts are also available for relative effect sizes as a non-parametric method \cite{Konietschke2012} with the package nparcomp \cite{Konietschke2015}. Thus, arbitrarily distributed endpoints with heterogeneous variances, which can be arbitrarily discrete from an asymptotic point of view, can be evaluated accordingly.

\begin{table}[ht]
\centering\small
\begin{tabular}{l|l|ll}
  \hline
 No& Type & Comparison & adjusted p-value \\ 
  \hline
1 & Males  &1:1 - 0 & 0.087 \\ 
  2 & &1:2 - 0 & 0.331 \\ 
  3 & &1:3 - 0 & 0.684 \\ 
  4 & &1:4 - 0 & 0.537 \\ \hline
  5 & Females&2:1 - 0 & 0.399 \\  
  6 & &2:2 - 0 & 0.176 \\ 
  7 & &2:3 - 0 & 0.040 \\ 
  8 & &2:4 - 0 & 0.037 \\ \hline
  9 & Pooled &p: 1 - 0 & 0.075 \\ 
  10 & & p: 2 - 0 & 0.049 \\ 
  11 & & p: 3 - 0 & 0.041 \\ 
  12 & & p: 4 - 0 & 0.022 \\ 
   \hline
\end{tabular}
\caption{Joint nonparametric Dunnett tests for males, females and pooled data in the IBS trial}
\label{tab:Egbertnon} 
\end{table}

Compared to the parametric analysis (see Table \ref{tab:Egbert}), the conclusions are similar, but the clearest effect here is at pooled consideration of comparing dose 4 with placebo (0).

\subsection{Comparing proportions by means of generalized linear models approach}
The rate of reocclusion were considered in a three-arm randomized clinical trial comparing aspirin and aspirin + dipyridamole with placebo  in 6 different sites (group G1 in \cite{Chootrakool2011}).	 
\begin{figure}[h]
	\centering
		\includegraphics[width=0.35\textwidth]{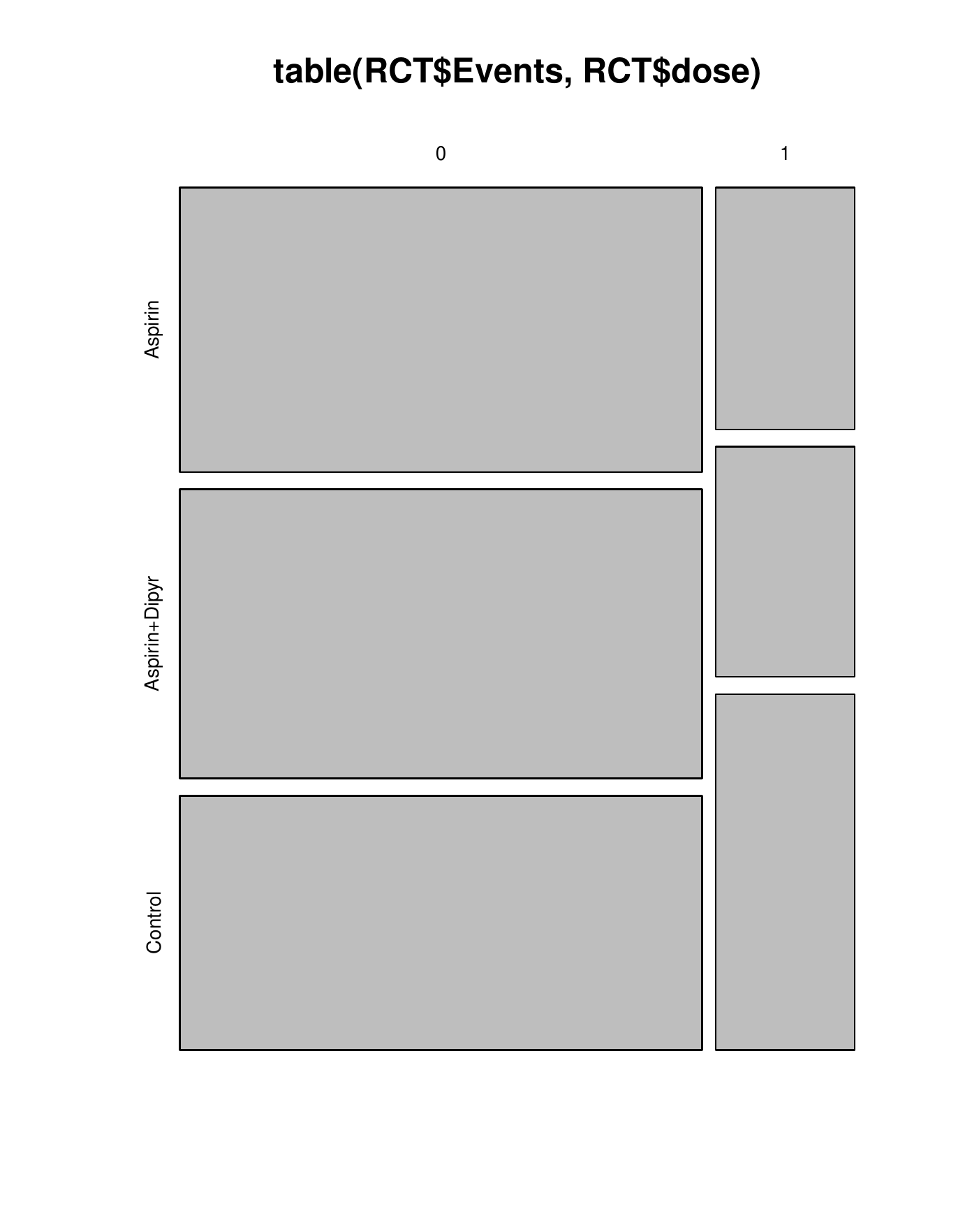}
		\includegraphics[width=0.35\textwidth]{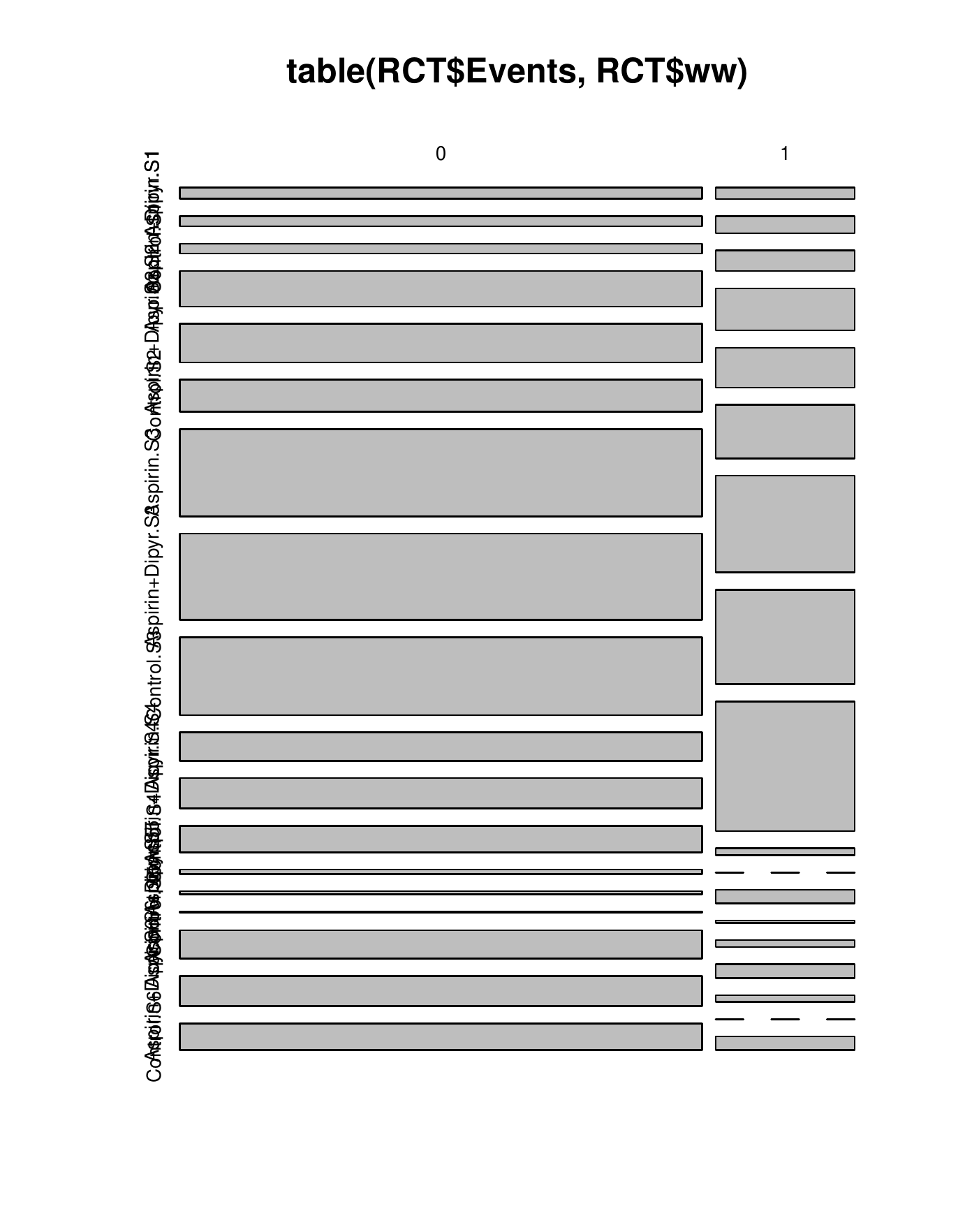}
	\caption{Mosaicplot reocclusion rates (left: global, right: per site)}
	\label{fig:MosaG}
\end{figure}

Adjusted p-values are reported in Table \ref{tab:glm}  for one-sided inference versus control using the add-two approximation for zero-control table data \cite{Agresti2011}:

\begin{table}[ht]
\centering\small
\begin{tabular}{rlrrrrr}
  \hline
No & Contrast & Site & estimate & se & t & p \\ 
  \hline
1 & Drug Aspirin - Control & S1 & -0.7 & 0.45 & -1.47 & 0.5730 \\ 
  2 & Drug Aspirin+Dipyr - Control & & -0.2 & 0.42 & -0.48 & 0.9833 \\ 
  3 & Drug Aspirin - Control & S2 & -0.3 & 0.25 & -1.33 & 0.6674 \\ 
  4 & Drug Aspirin+Dipyr - Control & & -0.5 & 0.26 & -1.82 & 0.3427 \\ 
  5 & Drug Aspirin - Control & S3 & -0.4 & 0.17 & -2.43 & 0.0921 \\ 
  6 & Drug Aspirin+Dipyr - Control & & -0.4 & 0.17 & -2.49 & 0.0768 \\ 
  7 & Drug Aspirin - Control & S4 & -0.7 & 0.49 & -1.39 & 0.6286 \\ 
  8 & Drug Aspirin+Dipyr - Control &  & -2.7 & 1.05 & -2.57 & 0.0645 \\ 
  9 & Drug Aspirin - Control & S5 & -2.4 & 0.80 & -2.97 & 0.0199 \\ 
  10 & Drug Aspirin+Dipyr - Control & & -1.2 & 0.69 & -1.77 & 0.3720 \\ 
  11 & Drug Aspirin - Control & S6 & -0.7 & 0.49 & -1.39 & 0.6284 \\ 
  12 & Drug Aspirin+Dipyr - Control &  & -2.7 & 1.05 & -2.57 & 0.0633 \\ \hline
  13 & Drug Aspirin - Control & Pooled & -1.3 & 0.22 & -5.94 & 0.0000 \\ 
  14 & Drug Aspirin+Dipyr - Control &  & -0.5 & 0.22 & -2.24 & 0.1474 \\ \hline\hline
	 & Drug Aspirin - Control & Global & -0.48 & 0.12 & -4.04 & 0.00005 \\ 
   & Drug Aspirin+Dipyr - Control & & -0.55 & 0.12 & -4.53 & 0.000006 \\    \hline
\end{tabular}
\caption{Joint glm-based Dunnett tests for per-site and pooled data}
\label{tab:glm}
\end{table}

\subsection{Considering variance heterogeneity}
Variance heterogeneity is quite common in bio-medical trials. Depending on the type of unbalancedness and related variance heterogeneity the Dunnett-type inference may be unacceptable liberal power-loss causing conservative or even can cause comparison-specific bias. Therefore, related adjustments by either using a sandwich estimator   \cite{Herberich2010}). in the linear model or adjusted degree of freedom(s) \cite{Hasler2008} \cite{Zhang2015} can be recommended. The above first example reveals clearly the advantage for such an approach

\section{Conclusions}
As an alternative to the usual pre-tests for interaction, a simultaneous procedure is proposed, which  realizes the inference of the primary factor at the respective level of the secondary factor as well as pooled over all its levels jointly.
This approach correctly controls for FWER, is somewhat conservative, but allows individual interpretation of the effects of potential interactions at the level of mean value comparisons. Both adjusted p-values and simultaneous confidence intervals (compatible with them)  are available. For designs with not too many factors and factor levels, this approach is relatively easy to realize with library(multcomp).\\
Another advantage of this approach is its flexibility for different types of variables, since it is based just on GLM's.

\footnotesize
\bibliographystyle{plain}

\begin{thebibliography}{10}

\bibitem{Dmitrienko2005}
Dmitrienko A., Molenberghs G., Chuang-Stein C., and Offen W.
\newblock Analysis of clinical trials using SAS: A practical guide.
\newblock {\em SAS Institute Inc: Cary, NC}, 2005.

\bibitem{Agresti2011}
A.~Agresti.
\newblock Score and pseudo-score confidence intervals for categorical data
  analysis.
\newblock {\em Statistics in Biopharmaceutical Research}, 3(2):163--172, May
  2011.

\bibitem{Biesheuvel2002}
E.~Biesheuvel and L.~A. Hothorn.
\newblock Many-to-one comparisons in stratified designs.
\newblock {\em Biometrical Journal}, 44(1):101--116, 2002.

\bibitem{Bretz2006}
F.~Bretz.
\newblock An extension of the Williams trend test to general unbalanced linear
  models.
\newblock {\em Computional Statistics and Data Analysis, vol. 50, no. 7, Art.
  no. 7, 2006.}, 2006.

\bibitem{Chen2007}
B.~Chen and J.~Shao.
\newblock Exact test for interaction in the two-way analysis of
  variance/covariance.
\newblock {\em Stat. Sinica}, 2007.

\bibitem{Cheung1996}
S.~H. Cheung and W.~S. Chan.
\newblock Simultaneous confidence intervals for pairwise multiple comparisons
  in a two-way unbalanced design.
\newblock {\em Biometrics}, 52(2):463--472, June 1996.

\bibitem{Chootrakool2011}
H.~Chootrakool, J.~Q. Shi, and R.~X. Yue.
\newblock Meta-analysis and sensitivity analysis for multi-arm trials with
  selection bias.
\newblock {\em Statistics in Medicine}, 30(11):1183--1198, May 2011.

\bibitem{Cribbie2016}
R.~A. Cribbie, C.~Ragoonanan, and A.~Counsell.
\newblock Testing for negligible interaction: A coherent and robust approach.
\newblock {\em British Journal of Mathematical and Statistical Psychology},
  69(2):159--174, May 2016.

\bibitem{DUNNETT1955}
C.~W. Dunnett.
\newblock A multiple comparison procedure for comparing several treatments with
  a control.
\newblock {\em Journal of the American Statistical Association},
  50(272):1096--1121, 1955.

\bibitem{Hasler2008}
M.~Hasler and L.~A. Hothorn.
\newblock Multiple contrast tests in the presence of heteroscedasticity.
\newblock {\em Biometrical Journal}, 50(5):793--800, October 2008.

\bibitem{Herberich2010}
E.~Herberich, J.~Sikorski, and T.~Hothorn.
\newblock A robust procedure for comparing multiple means under
  heteroscedasticity in unbalanced designs.
\newblock {\em PLOS One}, 5(3):e9788, March 2010.

\bibitem{Hothorn2020h}
L.A. Hothorn.
\newblock Statistical evaluation of in-vivo bioassays in regulatory toxicology
  considering males and females.
\newblock {\em arXiv}, 2020.

\bibitem{LH2022t}
L.A. Hothorn.
\newblock Hidden multiplicity in the analysis of variance (anova):multiple
  contrast tests as an alternative.
\newblock {\em bioRxiv https://doi.org/10.1101/2022.01.15.476452}, 2022.

\bibitem{Hothorn2006}
Ludwig~A. Hothorn.
\newblock {Multiple comparisons and multiple contrasts in randomized
  dose-response trials-confidence interval oriented approaches}.
\newblock {\em {J. Biopharm. Stat.}},
  {16}({5}):{711--731}, {2006}.

\bibitem{Hothorn2020t}
T.~Hothorn.
\newblock Additional multcomp examples to the CRAN package multcomp.
\newblock 2020.

\bibitem{Hothorn2008}
T.~Hothorn, F.~Bretz, and P.~Westfall.
\newblock Simultaneous inference in general parametric models.
\newblock {\em Biometrical J}, 50(3):346--363, 2008.

\bibitem{Tukey1953}
Tukey J.
\newblock The problem of multiple comparisons.

\bibitem{Kitsche2015}
A.~Kitsche and F.~Schaarschmidt.
\newblock Analysis of statistical interactions in factorial experiments.
\newblock {\em Journal of Agronomy and Crop Science}, 201(1):69--79, February
  2015.

\bibitem{Kluxen2020}
F.~M. Kluxen and L.~A. Hothorn.
\newblock Alternatives to statistical decision trees in regulatory
  (eco-)toxicological bioassays.
\newblock {\em Archives of Toxicology}, 2020.

\bibitem{Konietschke2012}
F.~Konietschke and L.~A. Hothorn.
\newblock Rank-based multiple test procedures and simultaneous confidence
  intervals.
\newblock {\em Electronic Journal of Statistics}, 6:738--759, 2012.

\bibitem{Konietschke2015}
F.~Konietschke, M.~Placzek, F.~Schaarschmidt, and L.~A. Hothorn.
\newblock nparcomp: An R software package for nonparametric multiple
  comparisons and simultaneous confidence intervals.
\newblock {\em Journal of Statistical Software}, 64(9), March 2015.

\bibitem{pallmann2016}
P.~Pallmann and L.~A. Hothorn.
\newblock Analysis of means: a generalized approach using R.
\newblock {\em Journal of Applied Statistics}, 43(8):1541--1560, June 2016.

\bibitem{Schaarschmidt2017}
F.~Schaarschmidt.
\newblock Multiple treatment comparisons in analysis of covariance with
  interaction sci for treatment covariate interaction.
\newblock {\em Statistical Methods and Applications}, 26(4):609--628, November
  2017.

\bibitem{WILLIAMS1971}
D.~A. Williams.
\newblock Test for differences between treatment means when several dose levels
  are compared with a zero dose control.
\newblock {\em Biometrics}, 27(1):103--, 1971.

\bibitem{wludyka2015}
P.~Wludyka.
\newblock Using ANOM slicing for multi-way models with significant
  interactions.
\newblock {\em J. Qualit. Technol.}, 2015.

\bibitem{Zhang2015}
G.~Y. Zhang.
\newblock Simultaneous confidence intervals for pairwise multiple comparisons
  in a two-way unbalanced design with unequal variances.
\newblock {\em Journal of Statistical Computation and Simulation},
  85(13):2727--2735, September 2015.

\bibitem{zhu2019}
J.~Zhu, K.~Lau, R.~Puschmann, R.~K. Harmel, Y.~J. Zhang, V.~Pries, P.~Gaugler,
  L.~Broger, A.~K. Dutta, H.~J. Jessen, G.~Schaaf, A.~R. Fernie, L.~A. Hothorn,
  D.~Fiedler, and M.~Hothorn.
\newblock Two bifunctional inositol pyrophosphate kinases/phosphatases control
  plant phosphate homeostasis.
\newblock {\em Elife}, 8:e43582, August 2019.

\end{thebibliography}

\par\medskip

\textbf{Acknowledgment:} My special thanks go to my son Prof. Dr. Torsten Hothorn (University of Zurich) for his inspiring lecture on the occasion of my retirement - exactly this was the starting point for the above approach.

\section{Appendix: R Code}
\subsection{Example: parametric test}
\scriptsize
\begin{verbatim}
library(DoseFinding)
data(IBScovars)
IBScovars$Dose<-as.factor(IBScovars$dose)

library(multcomp)
Mod1<-lm(resp~Dose+Gender, data=IBScovars)
Mod2<-lm(resp~Dose+Gender+Dose:Gender, data=IBScovars)
IBSm<-droplevels(IBScovars[IBScovars$gender==1, ])
IBSf<-droplevels(IBScovars[IBScovars$gender==2, ])
Mod3<-lm(resp~Dose, data=IBSm)
Mod4<-lm(resp~Dose, data=IBSf)
X1 <- glht(Mod1, linfct = mcp(Dose="Dunnett"),alternative="greater", vcov = sandwich)
pre<-anova(Mod2)[3,5]
X3 <- glht(Mod3, linfct = mcp(Dose="Dunnett"),alternative="greater", vcov = sandwich)
X4 <- glht(Mod4, linfct = mcp(Dose="Dunnett"),alternative="greater", vcov = sandwich)

Du <- contrMat(table(IBScovars$Dose), "Dunnett")
K1 <- cbind(Du, matrix(0, nrow = nrow(Du), ncol = ncol(Du)))
rownames(K1) <- paste(levels(IBScovars$Gender)[1], rownames(K1), sep = ":")
K2 <- cbind(matrix(0, nrow = nrow(Du), ncol = ncol(Du)), Du)
rownames(K2) <- paste(levels(IBScovars$Gender)[2], rownames(K2), sep = ":")
K <- rbind(K1, K2)
colnames(K) <- c(colnames(Du), colnames(Du))
IBScovars$ww <- with(IBScovars, interaction(Dose,Gender))
Mod5 <- lm(resp ~ ww - 1, data = IBScovars)

X5 <- glht(Mod5, linfct = K)
nn<-c(21,24,26,27,20,50,54,49,45,53)
w1<-nn[1]/(nn[1]+nn[6]); w6<-nn[6]/(nn[1]+nn[6])
w2<-nn[2]/(nn[2]+nn[7]); w7<-nn[7]/(nn[2]+nn[7])
w3<-nn[3]/(nn[3]+nn[8]); w8<-nn[8]/(nn[3]+nn[8])
w4<-nn[4]/(nn[4]+nn[9]); w9<-nn[9]/(nn[4]+nn[9])
w5<-nn[5]/(nn[5]+nn[10]); w10<-nn[10]/(nn[5]+nn[10])
K3<-rbind("p: 1 - 0" = c(-w1, w2, 0,   0,  0,  -w6,  w7, 0,   0,  0),
          "p: 2 - 0" = c(-w1, 0,  w3, 0,  0,  -w6,  0,  w8, 0,  0),
          "p: 3 - 0" = c(-w1, 0,  0,   w4,0,  -w6, 0,  0,   w9,0),
          "p: 4 - 0" = c(-w1, 0,  0,   0,  w5,-w6, 0,  0,   0,  w10))

K5<-rbind(K1,K2,K3)
library(sandwich)
X6<-glht(Mod5, linfct = K5, alternative="greater", vcov = sandwich)
library(ggplot2)
xx11<-fortify(summary(X1))
xx13<-fortify(summary(X3))
xx14<-fortify(summary(X4))
xx15<-fortify(summary(X5))
xx16<-fortify(summary(X6))
xx<-rbind(xx11[,c(1,5,6)],xx13[,c(1,5,6)],xx14[,c(1,5,6)],xx15[,c(1,5,6)],xx16[,c(1,5,6)])
library(xtable)
print(xtable(xx16[,c(1,6)], caption="c", digits=3))
\end{verbatim}
\normalsize

\subsection{Example: nonparametric test}
\tiny
\begin{verbatim}
library(nparcomp)
m4<-nparcomp(resp~ww, data=IBScovars, asy.method = "probit",
         type = "UserDefined", alternative = "greater",
         plot.simci = TRUE, info = FALSE,correlation=TRUE, contrast.matrix=K5)
S4<-summary(m4)
\end{verbatim}

\end{document}